\documentclass[showpacs,amsmath,amsart,twocolumn,showkeys,superscriptaddress]{revtex4-1}
\usepackage{dcolumn}    
\usepackage{ifpdf}
\usepackage{amssymb,lineno,amsfonts}
\usepackage{graphicx}   
\usepackage{bm}         
\usepackage{bbm}
\usepackage{mathrsfs}
\usepackage{upgreek}
\usepackage{mathtools}
\usepackage{epstopdf}
\usepackage{setspace}
\usepackage{hyperref}
\usepackage{float}
\usepackage{natbib}
\usepackage[usenames,dvipsnames]{xcolor}
\usepackage[matrix,frame,arrow]{xypic}
\usepackage{algorithmic}

\newcommand{\ket}[1]{\vert{#1}\rangle}

\newcommand{\proj}[1]{\vert{#1}\rangle\langle{#1}\vert}

\newcommand{\inpr}[2]{\langle{#1}\vert{#2}\rangle}
\newcommand{\expec}[1]{\langle{#1}\rangle}
\definecolor{med-blue}{RGB}{25,25,112}
\hypersetup{colorlinks, linkcolor={red},citecolor={blue}, urlcolor={MidnightBlue}}

\begin{document}
\title{Frequentist-approach inspired theory of quantum random phenomena\\predicts signaling}

\author{C. S. Sudheer Kumar} 
\affiliation{Department of Physics and NMR Research Center, Indian Institute of Science Education and Research, Pune 411 008, India}

\author{Anup Biswas} 
\affiliation{Department of Mathematics, Indian Institute of Science Education and Research, Pune 411 008, India}

\author{Aditi Sen(De)} 
\affiliation{Harish-Chandra Research Institute, HBNI, Chhatnag Road, Jhunsi, Allahabad 211 019, India}

\author{Ujjwal Sen} 
\affiliation{Harish-Chandra Research Institute, HBNI, Chhatnag Road, Jhunsi, Allahabad 211 019, India}

\begin{abstract}
Different ensembles of the same density matrix are indistinguishable within the modern Kolmogorov probability measure theory of quantum  random phenomena. 
We find that  changing the framework from the Kolmogorov one to a
frequentist-inspired theory of quantum random phenomena -- {\`a} la von Mises -- would lift the indistinguishability, and potentially cost us 
the  no-signaling principle (i.e., lead to superluminal communication).
We believe that this adds to the recent works on the search for a suitable representation of the state of a quantum system. 
While erstwhile arguments for potential modifications in the representation of the quantum state were restricted to possible variations in the formalism of the 
quantum theory, we indicate a possible fallout of altering the underlying theory of random processes.
\end{abstract}


\maketitle
\section{Introduction}
\noindent  Born's statistical interpretation of the state vector in quantum mechanics (QM)
and hence the density matrix description is based on 
Kolmogorov's modern axiomatic, probability-measure theoretic approach to 
random phenomena 
\cite{Prob_book_AlanGut,Prob_book_Chow,Sheldon_Ross_probability_book, Prob_measr_book, Prob_book_weigh_odd}. We refer to this as Kolmogorov QM (KQM) 
\cite{Peres_QM_book,QM_Cohen,Mathews-Venkatesan, QM_book_RShankar, QM_Griffiths}. 
The circularity in Kolmogorov's \textit{a priori} assumption of a constant value for the probability of a single random event and its subsequent justification via the strong law of 
large numbers (LLN), is well known \cite{Prob_book_Chow, circularLLNSpanos2013}.
It is to be noted that the convergence shown by the strong LLN 
is 
in terms of probability but not pointwise \cite{Prob_book_Chow, circularLLNSpanos2013}. This circularity might be a consequence of G{\"o}del's incompleteness theorem \cite{Emperor_new_mind_penrose, Goedel-incompleteness}.
The parallel and earlier approach by von Mises employs a limiting 
relative frequency definition of probability, which assumes existence of the limit \cite{vonMises_prob_book, vonMises_work_review, probInterprtn}, 
while it (the limit) does not exist in a strict mathematical sense 
\cite{Sheldon_Ross_probability_book, Prob_book_Chow, circularLLNSpanos2013}. 
Here we take an approach to quantum random phenomena which is inspired by the frequentist one, but different. We refer to it as 
frequentist-inspired QM (FQM). Conceptually, FQM is same as pathwise or model-free approach to stochastic processes in mathematical finance, wherein a probability measure is not assumed \textit{a priori} \cite{FollmerBook, pathwiseKARANDIKAR199511, pathwiseRamaConnt, pathwiseFinanceCandia, Hobson2011}.
We then show that such a frequentist-inspired approach leads to violation of the no-signaling principle \cite{No_communication_Peres, No_signal_in_time_Kofler, Bellvio4Popescu1994, Peres_QM_book} (i.e., leads to superluminal communication), by distinguishing between two different ensemble preparation procedures, 
which are indistinguishable in KQM, 
while still remaining within the Hilbert space 
formalism of quantum mechanics.

This work is organized as follows. In Sec. \ref{fqmintroduced}, we introduce the FQM and show that two ensembles described by the same density matrix can be distinguished via content-dependent relative fluctuations. In Sec. \ref{kqmwithfqm} we show how, for practical purposes, one can use KQM along with FQM in a consistent way. In Sec. \ref{signal}, we discuss the possibility of signaling within FQM. In Sec. \ref{Htheorem}, we discuss about the likely connection between Boltzmann's H-theorem and FQM. In Sec. \ref{Furtheraspects}, we briefly discuss about perfect anti-correlation of singlet even within FQM, and the case of finite number of trials. And we conclude in Sec. \ref{conclude}.

\section{A frequentist-inspired approach to quantum random phenomena}\label{fqmintroduced}
 Consider a random variable $X$ which is the outcome of projectively measuring 
\(|0\rangle\langle 0|\)
on $\ket{+}$ where $\ket{\pm}=(\ket{0}\pm\ket{1})/\sqrt{2}$, 
and $\ket{0}\) \((\ket{1})$ is the eigenstate of the Pauli-\(z\) observable, $\sigma_z$, with eigenvalue $+1\) \((-1)$. $X$ has the sample space $\{+1,0\}$. Assume that the measurement 
can be repeated indefinitely, under exactly the same conditions, on identical copies of $\ket{+}$, independently. KQM assumes, \emph{a priori}, a constant value for the probability 
of a single random event $X=+1$, based on the subjective notion of ``equally likely'' events, which is $P(X=+1)=1/2$ (Born's statistical interpretation of $\ket{+}$ \cite{QM_Griffiths}) 
\cite{Sheldon_Ross_probability_book,Prob_book_Chow,Prob_book_AlanGut, Prob_measr_book, Prob_book_weigh_odd}. 
However, note that the randomness may not be in the state $\ket{+}$. Indeed, it is possible to write down hidden variable theories of 
a single quantum spin-1/2 system that can predict, with certainty, the results of all measurements on it \cite{banshi-shune-aar-kaj-nai}. (Such 
hidden variables have never been observed in experiments.) Therefore, it may not be correct to characterize the random variable $X$ (i.e., assume that $P(X=+1)=1/2$) 
based solely on $\ket{+}$ (which may be equivalent to choosing it to be independent of something that causes randomness in measurement outcomes). According to de Finetti, constant objective probability does not exist \cite{Finetti_imprcProb, pathwiseFinanceCandia}. Hence it
may be worthwhile to keep the assumptions to the minimum possible (Ockham's razor \cite{OckhamRazorStanEncyclo}), and derive or obtain the rest of the structure or components experimentally 
(at least, at the conceptual level). In
FQM, we suppose that the objective limit-supremum of relative frequency (LRF) of the event $X=+1$, denoted as $F(X=+1)$ (this plays the role of $P(X=+1)$), 
is obtained \emph{a posteriori} via experiment as follows. 
Let $X_i$ be the outcome of the $i^{\mathrm{th}}$ 
trial of $X$. Then the number of $+1$ outcomes in $N$ independent trials of $X$ is given by 
$N_{+1}(X,N)=\sum_{i=1}^{N}X_i$. 
An operationally motivated 
definition of LRF of the event $X=+1$ is
\begin{eqnarray}
F(X=+1)=\limsup_{N\rightarrow\infty}\frac{N_{+1}(X,N)}{N}\nonumber\\
\equiv \lim_{N\rightarrow\infty}\bigg(\sup_{M\ge N}\frac{N_{+1}(X,M)}{M}\bigg)  
\coloneqq
  \frac{1}{2}+\kappa(X=+1)
\label{Assumptn}
\end{eqnarray}
\cite{RealAnalysis_basics_Houshang, RealanlysisRoyden, Math_analysis_book, Topology_Munkres, RealAnalysis_Apostl, Amritava-Gupta}, where $\kappa(X=+1)$ is a random variable which takes 
values in 
\([-\epsilon,\delta]
~(\epsilon>0,~\delta>0)$, depending on the outcomes in a given experiment. Note that in Eq. (\ref{Assumptn}), $1/2$ cannot be preferred over $1/2+c,|c|>0$, due to fundamental indeterminacy. Only relative fluctuation matters. (See Appendix \ref{halfnotpreoverhalfplusapp} for details.)
$\kappa(X=+1)$ represents an intrinsic or fundamental fluctuation in $F(X=+1)$. $\kappa(X=+1)$ is a consequence of Knightian type of `true' uncertainty \cite{KnightsBook, KnightUncertntyFinancePaper, Hobson2011, pathwiseFinanceCandia}. It is important to note that this 
fluctuation in $F(X=+1)$ is due to an intrinsic random
 nature of outcomes of the trials, 
and 
not due to varying conditions from one experiment to another, 
including 
imperfections in preparing a quantum state which are unavoidable 
in the real world. 
Similarly, we also define $F(X=0)=1/2+\kappa(X=0)$. Further, we define the limit-infimum of relative frequency of the event $X=+1$ as follows: $F'(X=+1)=1/2+\kappa'(X=+1)$. Note that as $F(X=+1)$ and $F(X=0)$ are independent, they need not sum to unity, unlike in $ F'(X=+1)+F(X=0)=1$. However for $N<\infty$, as there is no need of supremum and infimum, we have $F_N(X=+1)+F_N(X=0)=1$ where $\limsup_{N \to \infty}F_N(.)=F(.)$, and $\liminf_{N \to \infty}F_N(.)=F'(.)$. 
Note that $F(X=+1)$ is a random variable, whereas $P(X=+1)$ is a constant. 
It is important to note that $\lim_{N\rightarrow\infty}N_{+1}(X,N)/N$ cannot always converge 
pointwise (in event space) \cite{Math_analysis_book} to $1/2$, unlike, say, $\lim_{N\rightarrow\infty}1/N=0$ and 
$\lim_{N\rightarrow\infty}N_{+1}(X,N)/N^2\le \lim_{N\rightarrow\infty}1/N=0$ 
\cite{Prob_book_Chow, No_pointwise_convrg}. 
This is because $N_{+1}(X,N)$ is a random variable. 
The fundamental fluctuation in LRF can be considered as a \emph{resource} within the frequentist-inspired theory of quantum random phenomena, 
in particular, as we show now, for distinguishing between two different ensemble preparation procedures of 
the same density matrix.
This cannot be obtained 
within 
KQM due to \textit{a priori} assuming constant values for the corresponding probabilities.

\subsection{Distinguishing between two different ensemble preparation procedures for the same density matrix}\label{ensembldiscrm}
 Consider the two following preparation procedures.\\
\textbf{Procedure A:} In a trial of $X$, if the outcome is $+1$ ($0$), then Alice prepares a 
qubit in the state $\ket{0}$ ($\ket{1}$). She repeats the preceding step $\mathcal{M}$ times independently. She gives this bunch -- 
call it $\mathcal{E}_\mathrm{A}$ -- of $\mathcal{M}$ qubits to Bob.\\
\textbf{Procedure B:} This is the same as procedure A, except that $\ket{0}$ ($\ket{1}$) is replaced by $\ket{+}\) \((\ket{-})$.
Again, Alice hands over 
this bunch -- call it $\mathcal{E}_\mathrm{B}$ -- of $\mathcal{M}$ qubits to Bob.\\ 
Bob is aware of the two preparation procedures but unaware of the outcomes of trials of $X$. 
Further, Bob is allowed to choose the number \(\mathcal{M}\) as large as he decides,
carry out any unitary operation on the states, and measure any observable. 
The question is whether Bob can 
distinguish between the procedures A and B. The answer, within standard KQM, is in the negative, as the density matrix corresponding to both the 
procedures is the same, viz., \((\frac{1}{2}|0\rangle\langle 0|+\frac{1}{2}|1\rangle\langle 1|)^{\otimes\mathcal{M}}\). We now consider the solution within FQM.

Instead of representing the states of the bunches, \(\mathcal{E}_\mathrm{A}\) and \(\mathcal{E}_\mathrm{B}\), in terms of density matrices, one may choose to represent 
them path by path as
\begin{eqnarray}
\ket{\psi_j^\mathrm{A}}&=&\bigotimes_{i=1}^{\mathcal{M}}\ket{X_i\oplus 1}, \nonumber \\
\ket{\psi_j^\mathrm{B}}&=&\bigotimes_{i=1}^{\mathcal{M}}\ket{Z_i},
\label{psiAdefind}
\end{eqnarray}
where $\oplus$ is addition modulo 2,
$Z_i=+(-)$ if $X_i=+1(0)$, and $j\in\{1,2,...,2^{\mathcal{M}}\}$ \cite{ensembl_discrm_NMR_Found_phy}. Particles in a bunch are noninteracting.
Also, as particles in a bunch are distinguishable, Bob can ignore symmetrizing or 
anti-symmetrizing the total wave function representing the state of $\mathcal{E}_{\mathrm{A/B}}$ \cite{QM_book_RShankar, ensembl_discrm_dEspanag_PLA}.

The state $\ket{\psi^\mathrm{A(B)}_j}$ has all the information which Bob has about the given $\mathcal{E}_{\mathrm{A(B)}}$. 
It may be noted that $|\inpr{\psi^\mathrm{A}_j}{\psi^\mathrm{B}_k}|=\frac{1}{2^{\mathcal{M}/2}}\ne 1,\forall j,k$. 
See
\cite{Popescu_conjctr, AOC_lindep} 
in this respect. It may also be interesting to consider Refs. \cite{superactvtn_peres,superactvtn_Sen,superactivtn_palazuola} and references therein, where ``superactivation of nonlocality'' is considered within KQM. 

Bob applies 
\begin{equation}
R_x(X^\Theta)=\exp(-iX^\Theta\sigma_x/2)
\label{rak-khas-gan} 
\end{equation}
 to 
each of 
the qubits,
where $X^\Theta$ is a random variable which outputs $\theta_i$ with LRF $F(X^\Theta=\theta_i)=1/2+\kappa(X^\Theta=\theta_i),i=1,2$. Then he measures $\sigma_z$ on the qubit state.

Suppose, unknown to Bob, the bunch that he obtained was created by procedure A. 
Now, $R_x(X^\Theta=\theta_n)\ket{0}=\ket{\theta_n,-\pi/2}\), and \(R_x(X^\Theta=\theta_n)\ket{1}=-i\ket{\pi-\theta_n,\pi/2}\), for \(n=1,2$, 
 where $\ket{\theta,\phi}=\cos\frac{\theta}{2}\ket{0}+e^{i\phi}\sin\frac{\theta}{2}\ket{1}$ in the usual Bloch sphere representation. Let $X^{\theta}$ be 
the outcome of measuring $\sigma_z$ on $\ket{\theta,\phi}$. Then,
\begin{eqnarray}
F(X^\theta=+1)=\cos^2(\theta/2)+\kappa(X^\theta=+1),
\label{Bornrulemodfied}
\end{eqnarray}
which is the \textit{modified Born's statistical interpretation of} $\ket{\theta,\phi}$. Note that here we have assumed that the fluctuation term i.e., $\kappa(.)$ will depend on content/state i.e, $\theta$. See Appendix \ref{kappadependsonthetaapp} for its justification.
And $F(X^\theta=-1)=\sin^2(\theta/2)+\kappa(X^\theta=-1)$, $\theta\ne 0,\pi$. Define sample mean as
            \begin{eqnarray}
             S(\mathrm{A},M)=\frac{1}{M}\sum_{i=1}^{M}X_i^{\theta},
             \label{samplemeanA}
            \end{eqnarray}
where $\mathrm{A}=\{X,X^\Theta,X^{\theta_1},X^{\theta_2},X^{\pi-\theta_1},X^{\pi-\theta_2}\}$, 
$X^\theta_i\in\{X^{\theta_1}_i,X^{\theta_2}_i,X^{\pi-\theta_1}_i,X^{\pi-\theta_2}_i\}$. Let $M=1$. Then
\begin{eqnarray}
F(S(\mathrm{A},M=1)=+1)=\limsup_{N\rightarrow\infty}\frac{N_{+1}(S(\mathrm{A},M=1),N)}{N}.\nonumber \\
\label{FofSM1gen}
\end{eqnarray}
 The LRF, as defined in Eq. (\ref{Assumptn}) or Eq. (\ref{FofSM1gen}), is the only experimental or operational way to characterize or gain information about a given random variable. Hence, any function we define should be expressible in terms of LRFs. The sample mean, as defined in Eq. (\ref{samplemeanA}) is one such function, as it can be rewritten as $\limsup_{N\rightarrow\infty}S(\mathrm{A},N)=2F(S(\mathrm{A},M=1)=+1)-1$. It is the average of Bob's final $\sigma_{z}$ measurement outcomes. (See Appendix \ref{physclsignificofSapp} for details.)

We first consider the situation where $\theta_2=\theta_1$. 
In this case,  
 $N_{+1}(S(\mathrm{A},M=1),N)=N_{+1}(X_1^{\theta_1},N_{+1}(X_1,N))+N_{+1}(X_1^{\pi-\theta_1},N_0(X_1,N))$, where $N_0(X_1,N)=N-N_{+1}(X_1,N)$. 
We have 
\begin{eqnarray}
&&\limsup_{N\rightarrow\infty}\frac{N_{+1}(X_1^{\theta_1},N_{+1}(X_1,N))}{N_{+1}(X_1,N)}\frac{N_{+1}(X_1,N)}{N}\nonumber\\
&\le&\limsup_{N\rightarrow\infty}\frac{N_{+1}(X_1^{\theta_1},N_{+1}(X_1,N))}{N_{+1}(X_1,N)}\limsup_{N\rightarrow\infty}\frac{N_{+1}(X_1,N)}{N}\nonumber\\
&=&(\cos^2\frac{\theta_1}{2}+\kappa(X_1^{\theta_1}=+1,+1(X_1)))(\frac{1}{2}+\kappa(X_1=+1)),~~~~
\label{N1byN}
\end{eqnarray}
for $N_{+1}(X_1,N\rightarrow\infty)>0$ 
 \cite{RealAnalysis_basics_Houshang, Realanlysiskaczor}. In Eq. (\ref{FofSM1gen}), using $\limsup_{N \to \infty}(x_N+y_N)\le\limsup_{N \to \infty}x_N+\limsup_{N \to \infty}y_N$ where $\{x_N\}$, \(\{y_N\}\) are sequences of real numbers \cite{RealanlysisRoyden}, and then substituting ineq. (\ref{N1byN}) 
and a similar result for $\limsup_{N\rightarrow\infty}N_{+1}(X_1^{\pi-\theta_1},N_0(X_1,N))/N$, we get 
\begin{eqnarray}
&&F(S(\mathrm{A},M=1)=+1) \le \frac{1}{2} \nonumber \\
&&+\kappa(X_1=+1)\left(\cos^2(\theta_1/2)+\kappa(X_1^{\theta_1}=+1,+1(X_1))\right)\nonumber\\
&& +\kappa(X_1=0)\left(\sin^2(\theta_1/2)+\kappa(X_1^{\pi-\theta_1}=+1,0(X_1))\right)\nonumber\\
&&+ \frac{1}{2}\left(\kappa(X_1^{\theta_1}=+1,+1(X_1))+\kappa(X_1^{\pi-\theta_1}=+1,0(X_1))\right).~~~~
\label{FofSM1}
\end{eqnarray}
(See Appendix \ref{evallimsupapp} for details.)
And $F(S(\mathrm{A},M=1)=-1)$ will have a similar expression. 

We note here that if we modify the KQM initial density matrix 
into $\rho^{\mathrm{A}}=(1/2+\kappa(X=+1))\proj{0}+(1/2+\kappa(X=0))\proj{1}$, then one can easily verify that 
$R_x(X^\Theta=\theta_1)\rho^{\mathrm{A}}R_x(X^\Theta=\theta_1)^\dagger$ along with the usual KQM Born rule for the subsequent \(\sigma_z\)-measurement 
do not reproduce the required result consistent with  ineq. (\ref{FofSM1}). 

Next suppose that the bunch of \(\mathcal{M}\) states that Bob obtained from Alice was prepared by procedure B. As before, Bob is oblivious of this choice of 
Alice. We have $R_x(X^\Theta=\theta_n)\ket{\pm}=e^{\mp i\theta_n/2}\ket{\pm}$, $n=1,2$.  Therefore,
\begin{eqnarray}
S(\mathrm{B},M) =\frac{1}{M}\sum_{i=1}^{M}X_i^{\pi/2},
\label{samplemean_B}
\end{eqnarray}
where $\mathrm{B}=\{X^{\pi/2}\}$. Then 
\begin{eqnarray}
&&F(S(\mathrm{B},M=1)=+1) \nonumber \\
&&=\limsup_{N\rightarrow\infty}\frac{N_{+1}(S(\mathrm{B},M=1),N)}{N} \nonumber \\
&&=\frac{1}{2}+\kappa(X_1=+1)\nonumber\\
&&=\frac{1}{2}+\kappa(X^\Theta_1=\theta_1),
\label{PofS_B}
\end{eqnarray}
since
$X^{\pi/2},X$, and $X^\Theta$ differ only in the value assigned to their outcomes. And
$F(S(\mathrm{B},M=1)=-1)=1/2+\kappa(X_1=0)$.

For $\theta_1=0,\pi/2$, ineq. (\ref{FofSM1}) reduces to 
$F(S(\mathrm{A},M=1)=+1)=1/2+\kappa(X_1=+1)$, because \(N_{+1}(X_1^{\theta_1=0},N_{+1}(X_1,N))=N_{+1}(X_1,N)\). (See Appendix \ref{theta1eqtheta2eq0app} for details.) 
However, in general the 
fluctuation of $F(S(\mathrm{A},M=1)=+1)$ (ineq. (\ref{FofSM1})) relative to that of $F(S(\mathrm{B},M=1)=+1)$ (Eq. (\ref{PofS_B})) is different. This is because, fluctuation of $\kappa(Y=y,Z)$ depends on both random variables $Y$ and $Z$. And the expressions (\ref{FofSM1}) and (\ref{PofS_B}) are different functions of $\kappa(...)$'s and it is impossible to reduce ineq. (\ref{FofSM1}) into Eq. (\ref{PofS_B}). (Also see Fig. \ref{plotfluctn}.) This is a necessary and sufficient condition for the discrimination. (See Appendix \ref{observbltyofcontentdepflucapp} for further justification.)
Assuming that there are no further physical restrictions on the observability of the fluctuations, 
we have therefore shown that our frequentist-inspired approach distinguishes equal density matrices. 

Further it is important to note that as relative fluctuation (which do not require quantitatively precise prediction) is sufficient for discriminating between the two preparation procedures, it is not really necessary to use KQM (which gives quantitatively precise prediction) even in the later stages of the calculations as done in Appendix \ref{normaldistrbforfappapp}. Hence the discrimination between the two preparation procedures is predicted completely within FQM.

\section{Using KQM along with FQM in a consistent way}\label{kqmwithfqm}
 If we use only FQM (KQM) then we obtain fundamentally correct (incorrect) but quantitatively imprecise (precise) predictions. KQM's prediction is fundamentally incorrect due to the unjustifiable nature of \textit{a priori} assumed probability measure. FQM's prediction is quantitatively imprecise due to the fundamental fluctuation associated with $\kappa(...)$'s. Hence, we should use KQM along with FQM, but in a  consistent way, to make physically correct as well as quantitatively precise  predictions. 
In fact, FQM or the ``pathwise'' approach is already being used 
(without it being stressed)
in quantum teleportation \cite{Teleport_benet_orginal, quant_info_neilson_chuang}, approximate quantum cloning \cite{noclone_buzek5by6}, the Bennett-Brassard 1984 quantum cryptography protocol \cite{BB84_original_paper}, \cite{quant_info_neilson_chuang}, discriminating between linearly independent \cite{USD_chefls_linind} and dependent \cite{CTC_Brunn_breakbb84} state vectors. In \cite{Teleport_benet_orginal}, \cite{noclone_buzek5by6}, \cite{BB84_original_paper}, and \cite{USD_chefls_linind,CTC_Brunn_breakbb84}, the authors consider the unknown states 
($\ket{\psi}$) 
to be teleported, cloned, cryptographed, and discriminated, respectively, within a path by path approach, and without assuming \textit{a priori} a probability measure for $Y$, where a single copy of $\ket{\psi}$ has been prepared according to the outcome of a trial of a random variable $Y$. However they assume \textit{a priori} a probability measure for other random variables.
We note 
here 
that 
in pathwise approach of mathematical finance \cite{FollmerBook}, probability measure is 
also
brought in at a later stage of the analysis to study the interplay between all paths of a given stochastic process. We note that the two notions, \textit{viz.}, assuming \textit{a priori} a probability measure $p_{\ket{\psi(y)}}$ for $Y$ and path by path consideration of $\ket{\psi}$'s, cannot exist simultaneously. If we assume \textit{a priori} a probability measure then we are forced to consider the average mathematical state, $\int_{\ket{\psi}} dp_{\ket{\psi}}\proj{\psi}$, 
instead of 
the actual physical states, $\ket{\psi}$
(see \cite{CTC_Benett} in this regard).


\begin{figure}
  	\centering
  	\includegraphics[width=9cm,clip = true,trim = 0.0cm 2cm 0.0cm 2cm]{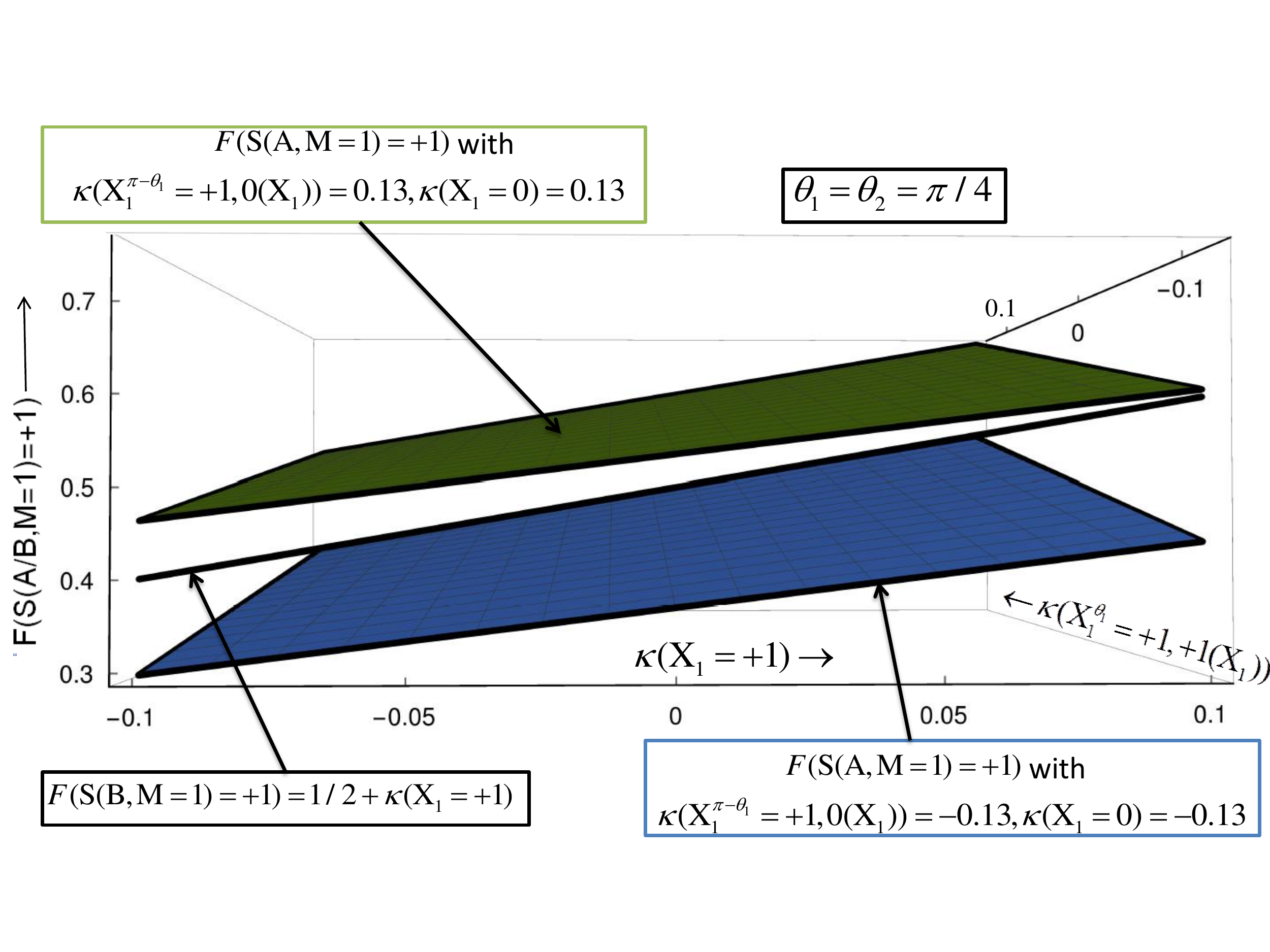}
\caption{
Comparing the frequentist predictions for two preparation procedures A and B. We consider here the $\theta_1=\theta_2$ case.   
We wish to compare \(F(S(\mathrm{A},M=1)=+1)\) with \(F(S(\mathrm{B},M=1)=+1)\). We set \(\theta_1 = \theta_2 = \pi/4\). 
 We have four independent random variables 
viz., $\kappa(X_1=+1),\kappa(X_1=0)\), \(\kappa(X_1^{\theta_1}=+1,+1(X_1))$, and $\kappa(X_1^{\pi-\theta_1}=+1,0(X_1))$. We present a ``front view'' i.e., looking along the normal to the $(\kappa(X_1=+1),F(S(\mathrm{A/B},M=1)=+1))$-plane. Hence, plot for \(F(S(\mathrm{B},M=1)=+1)\) is the simple black straight line. However the bounds of $F(S(\mathrm{A},M=1)=+1)$ are surfaces in the corresponding five-dimensional space. 
For given values of $\kappa(X_1^{\pi-\theta_1}=+1,0(X_1))$ and $\kappa(X_1=0)$, the same are surfaces in the corresponding three-dimensional space. $\kappa(X_1^{\pi-\theta_1}=+1,0(X_1))$ and $\kappa(X_1=0)$
can take both positive and negative values. Consider, first, an exemplary situation where $\kappa(X_1^{\pi-\theta_1}=+1,0(X_1))=-0.13$ and $\kappa(X_1=0)=-0.13$. 
This leads to the blue surface at the bottom for the bound of \(F(S(\mathrm{A},M=1)=+1)\) in ineq. (\ref{FofSM1}).
\(F(S(\mathrm{A},M=1)=+1|\kappa(X_1^{\pi-\theta_1}=+1,0(X_1)) = -0.13,\kappa(X_1=0)=-0.13)\) can only be below the blue surface, and so must be different from \(F(S(\mathrm{B},M=1)=+1)\).
The green surface, that is at the top for most of the considered region on the $(\kappa(X_1=+1), \kappa(X_1^{\theta_1}=+1,+1(X_1)))$-plane, 
 is the plot for the bound of $F(S(\mathrm{A},M=1)=+1)\) in ineq. (\ref{FofSM1})  
 with $\kappa(X_1^{\pi-\theta_1}=+1,0(X_1)) = 0.13$ and $\kappa(X_1=0)=0.13$. This time, 
\(F(S(\mathrm{A},M=1)=+1|\kappa(X_1^{\pi-\theta_1}=+1,0(X_1)) = 0.13,\kappa(X_1=0)=0.13)\) can only be below 
the green surface, and again there are regions where it is different from $F(S(\mathrm{B},M=1)=+1)\).
Hence in procedure A, there are points corresponding to LRF 
which are above, as well as below that corresponding to end points of the line segment $F(S(\mathrm{B},M=1)=+1)=1/2+\kappa(X_1=+1)$.
The fluctuation of LRF, around 1/2, will therefore be different in the two procedures. 
For ease of plotting,
we have taken $|\kappa(\cdots)|$'s to be large.
All quantities are dimensionless. Note that the surfaces in the above figure gives only the upper bounds. To know the corresponding lower bounds, we need to evaluate limit infimum. (See Appendix \ref{evalliminfapp} for details.)
  	}
\label{plotfluctn}
\end{figure}   
 
\section{Signaling}\label{signal}
The distinguishing protocol discussed above can be used to provide instantaneous transfer of information between two separated locations.
See \cite{Popescunonlocalbeyndqmreviewnpj, Bellvio4Popescu1994, QKDmongmyPawlosky, PTsym_vio_nosignal} in this respect. 
Let Alice and Bob share $\mathcal{M}$ singlets $\ket{S_0}=(\ket{01}-\ket{10})/\sqrt{2}$, and be space-like separated. If Alice measures $\sigma_z(\sigma_x)$ on her qubits, 
then on Bob's side $\mathcal{E}_\mathrm{A(B)}$ is produced. As Bob can distinguish (at least in principle) between 
$\mathcal{E}_\mathrm{A}$ and $\mathcal{E}_\mathrm{B}$, he can know Alice's measurement choice superluminally. Note that we are not using nonlinear evolution to achieve signaling, like in \cite{CTC_Brunn_breakbb84,PTsym_vio_nosignal}.

\section{Connection to H-theorem}\label{Htheorem}
 The Boltzmann entropy of a non-equilibrium physical system, increases with time, as per the H-theorem. However, the Gibbs-von Neumann entropy of the same system, is constant in time (consequence of Liouville's theorem). The two definitions of entropy agree in equilibrium systems \cite{Gibs_Boltz_entrpy_FQM}. The Boltzmann entropy is defined within an approach where we consider the actual state of the given physical system (i.e., path by path approach \cite{FollmerBook, pathwiseKARANDIKAR199511, pathwiseRamaConnt, pathwiseFinanceCandia, Hobson2011}) without assuming, \textit{a priori}, a probability measure. Whereas, the Gibbs-von Neumann entropy is based on the density matrix approach wherein we assume, \textit{a priori}, a probability measure to obtain the average state of the system under consideration. The proof of the H-theorem depends on the definition of Boltzmann entropy, and crucially uses the hypothesis of ``molecular chaos'' or ``past-hypothesis'' or ``typicality'' \cite{Gibs_Boltz_entrpy_FQM, StatMechHuang, UniasPresrvEntropyUSen}, along with the Hamiltonian dynamics, while the constancy of the Gibbs-von Neumann entropy uses the Hamiltonian dynamics only. It seems that the additional assumption akin to molecular chaos cannot be employed within the density matrix formalism of state description. See \cite{Gibs_Boltz_entrpy_FQM, Kac_markovJumpBoltzGibsEntrpy} in this regard. Assuming that to be true, this implies that averaging via a probability measure to obtain a density matrix, used in the Gibbs-von Neumann entropy, erases some information relevant to the dynamics of non-equilibrium systems.

\section{Further aspects}\label{Furtheraspects}
 Consider $S(\sigma_{z}^\mathcal{A}\sigma_{z}^\mathcal{B},M)=(1/M)\sum_{i=1}^{M}\sigma_{zi}^\mathcal{A}\sigma_{zi}^\mathcal{B}$ where the random variable $\sigma_{zi}^\mathcal{A(B)}$ is the outcome of 
Alice (Bob) measuring $\sigma_z$ on her (his) $i^\mathrm{th}$ qubit in the state $\alpha\ket{01}+\beta\ket{10},|\alpha|^2+|\beta|^2=1$. Then in FQM, 
one can easily show that $\lim_{N\rightarrow\infty}S(\sigma_{z}^\mathcal{A}\sigma_{z}^\mathcal{B},N)=-1$.
(See Appendix \ref{perfectanticorsingltapp} for details.) 
Hence, even though one may feel that 
the randomness of $\kappa(\cdots)$ terms will get canceled by an extra randomness in the anti-correlation of the singlet and prevent signaling, such a thing 
does not happen, simply because such an extra randomness does not exist. Further, one may also feel that the randomness of $\kappa(\cdots)$ terms will get 
constrained by constraining the extra randomness in the anti-correlation of the singlet. This also does not happen for the same reason. 

For $1\ll N<\infty$, we obtain expressions which are same as the expressions (\ref{FofSM1}), (\ref{PofS_B}), but with $\kappa(\cdots)$'s replaced by the corresponding $\kappa_N(\cdots)$'s (which represent fluctuation corresponding to $1\ll N<\infty$ such that $\limsup_{N\rightarrow\infty}\kappa_N(\cdots)=\kappa(\cdots)$), inequalities replaced by equalities,  $\kappa_N(X=+1)+\kappa_N(X=0)=0$ and with similar constraint for other $\kappa_N(\cdots)$ terms. This is because, when we take the limit $N\rightarrow\infty$, it turns out that the limit may not exist. Hence we have to consider limit supremum or limit infimum which always exists, and they give rise to inequalities.  
(See Appendix \ref{finiteNcaseapp} for details.) 
Hence Bob can distinguish even when $1\ll N<\infty$.

The case when \(\theta_1 \ne \theta_2\) and the concept of using KQM in the later stages of calculations for practical purposes are considered in the Appendix \ref{theta1neqtheta2caseapp} and Appendix \ref{normaldistrbforfappapp} respectively. 


Note that if we set $\kappa(\cdots)$'s to $0$ in expressions (\ref{FofSM1}), (\ref{PofS_B}), 
we obtain the numerical values corresponding to the predictions of KQM. 
In this sense, KQM can be seen as a special case of FQM.


\section{Conclusion} \label{conclude}
In summary, we found that a frequentist-inspired theory of quantum random phenomena leads to distinguishing between different ensembles of the same density matrix, 
which in turn leads to signaling (i.e., superluminal communication). This may be seen in the light of previous comments about the possible incompleteness of the density matrix representation, within 
modern Kolmogorov probability measure theory of quantum random phenomena,
of a situation (state) of a physical system in Refs. \cite{Penrose_large_small_mind, ensembl_discrm_NMR_Found_phy, 
ensembl_discrm_dEspanag_PLA, YouQu17, Popescu_conjctr, Bell_spek_unspek_book, CTC_cgargeBenett_Cavalcanti}. 
To our knowledge, preceding discussions on possible modifications of the density matrix representation confined themselves to revisions of the 
description of the state within the Hilbert space formalism of quantum mechanics. We showed that remaining within the Hilbert space formalism but 
looking out for possible implications of variations of the underlying theory of random processes may cost us the no-signaling principle.

\section*{Acknowledgments}
We thank Anjusha V. S., Aravinda
S., Rupak Bhattacharya, Udaysinh T. Bhosale, Sreetama
Das, Deepak Dhar, Dipankar Home, Aravind Iyer, H. S.
Karthik, Deepak Khurana, Pieter Kok, Manjunath Krishnapur,
V. R. Krithika, T. S. Mahesh, Masanao Ozawa,
Soham Pal, Apoorva D. Patel, Arun K. Pati, A. K.
Rajagopal, M. S. Santhanam, Abhishek Shukla, Mohd.
Asad Siddiqui, R. Srikanth, Chirag Srivastava, Dieter
Suter, Govind Unnikrishnan, A. R. Usha Devi, Lev Vaidman,
and Marek {\.Z}ukowski for useful discussions.

\appendix
 \section{In $F(X=+1)$, $1/2$ cannot be preferred over $1/2+c$}\label{halfnotpreoverhalfplusapp}
 In Eq. (\ref{Assumptn}), choosing 1/2 is motivated/guided by the following factors: Experimental observation (i.e., stabilization of relative frequency \cite{Prob_book_AlanGut} somewhere around 1/2), symmetry i.e., $\ket{+}$ is an equal superposition of the two eigenvectors of the observable being measured (i.e., $\proj{0}$), and convenience i.e, 1/2 is the square of the Fourier coefficient or amplitude in $\ket{+}$. However from foundational point of view, these are not compelling and sufficient reasons to prefer 1/2 over $1/2+c,|c|>0$. (Note that if we choose $1/2+c$ then we will not recover KQM from FQM by setting $\kappa(.,.)$ terms to zero. But that is okay because any way in KQM \textit{a priori} probability 1/2 is not justifiable physically. Then there is no compelling reason for not to choose $1/2+c$ (instead of 1/2) as \textit{a priori} probability.) The fact that $F(X=+1)$ cannot always converge pointwise to 1/2 proves that even $F(X=+1)$ has intrinsic fluctuation. Further even if we repeat infinitely many times the experiment involving $N\to\infty$ number of trials of $X$, still $F(X=+1)$ may not always fluctuate symmetrically about 1/2. This is more appealing in case of $F(X^{\theta\ne\pi/2}=+1)$ where $X^\theta$ is defined in the text preceding Eq. (\ref{Bornrulemodfied}). This is due to fundamental uncertainty/indeterminacy arising due to intrinsic randomness in measurement outcomes. If $F(X=+1)$ would always fluctuate symmetrically about 1/2 then that would contradict the very meaning, nature, and definition of random phenomena. What really matters and one can talk of is the relative fluctuation i.e., fluctuation of $F(X^{\theta=\pi/2}=+1)$ will be different compared to that of $F(X^{\theta\ne\pi/2}=+1)$. This is unlike in KQM wherein one can talk of absolute fluctuation due to the presence of quantitatively precise probability measure. 
 \\Of course we can absorb $c$ into $\kappa(X=+1)$. But here we are trying to argue that $F(X=+1)$ may not always fluctuate symmetrically about 1/2. And hence there is no compelling reason to prefer 1/2 over $1/2+c$.

\section{Justification of the assumption that $\kappa(X^\theta=+1)$ depends on $\theta$}\label{kappadependsonthetaapp}
The fact that $F(X^\theta=+1)$ cannot always converge pointwise to $\cos^2(\theta/2)$ proves the existence of fluctuation term i.e., $\kappa(.)$, but it does not say if $\kappa(.)$ depends on $\theta$ or not. Hence in Eq. (\ref{Bornrulemodfied}) we have implicitly assumed that $\kappa(.)$ will depend on $\theta$. This can be justified as follows. KQM predicts that variance (which is a measure of fluctuation), $\mathrm{Var}(X^\theta)=\expec{{X^{\theta}}^2}-\expec{X^\theta}^2=\sin^2\theta$. This has been tested experimentally to a good extent. Hence from this we can deduce that fluctuation will be maximum for $\theta=\pi/2$ and fluctuation gradually decreases as $\theta$  either decreases to 0 or increases to $\pi$. Hence it is an experimental fact that fluctuation will depend on content/state i.e., $\theta$. This justifies the assumption that fluctuation of $F(X^\theta=+1)$ (and hence $\kappa(X^\theta=+1)$) will depend on $\theta$.

\section{Physical meaning and significance of sample mean}\label{physclsignificofSapp}
Consider sample means $S(\mathrm{A},M)$, $S(\mathrm{B},M)$ as defined in Eqs. (\ref{samplemeanA}, \ref{samplemean_B}) respectively. They are the average of final (i.e., after applying $R_x(X^\Theta)$) $\sigma_{z}$ measurement (carried out by Bob) outcomes $X^\theta_i$'s. In procedure A, $X^\theta_i\in\{X^{\theta_1}_i,X^{\theta_2}_i,X^{\pi-\theta_1}_i,X^{\pi-\theta_2}_i\}$. Whereas in procedure B, $X^\theta_i\in\{X^{\pi/2}_i\}$ (because in procedure B, with respect to $\sigma_{z}$ measurement outcomes, the states $\ket{+}$ and $\ket{-}$ are equivalent). For the sake of ease, let us consider $S(\mathrm{B},M)$ (because in procedure B, the set to which $X^\theta_i$ belongs to, has only one element).
We have defined
\begin{eqnarray}
S(\mathrm{B},M) =\frac{1}{M}\sum_{i=1}^{M}X_i^{\pi/2}.\nonumber
\end{eqnarray}
$\Rightarrow S(\mathrm{B},M=1) =X_1^{\pi/2}$. Hence we can rewrite
\begin{eqnarray}
S(\mathrm{B},N) =\frac{1}{N}(N_{+1}(S(\mathrm{B},M=1),N)\nonumber\\
-N_{-1}(S(\mathrm{B},M=1),N))\nonumber\\
=\frac{1}{N}(2N_{+1}(S(\mathrm{B},M=1),N)-N)\nonumber\\
(\because N_{+1}(S(\mathrm{B},M=1),N)+N_{-1}(S(\mathrm{B},M=1),N)=N)\nonumber\\
=2F_N(S(\mathrm{B},M=1)=+1)-1=2F_N(X^{\pi/2}_1=+1)-1.\nonumber
\end{eqnarray}
 Similarly one can obtain $S(\mathrm{A},N)=2F_N(S(\mathrm{A},M=1)=+1)-1$. Sample means are used to study the relative fluctuation in the two procedures A and B.

\section{Evaluating $\limsup_{N\rightarrow\infty}\frac{N_{+1}(X_1^{\pi-\theta_1},N_0(X_1,N))}{N}$}\label{evallimsupapp}
If $\{a_N\}$ and $\{b_N\}$ are sequences of non-negative numbers, then
\begin{eqnarray}
\liminf_{N\rightarrow\infty}a_N\liminf_{N\rightarrow\infty}b_N\le\liminf_{N\rightarrow\infty}(a_Nb_N)\nonumber\\
\le \liminf_{N\rightarrow\infty}a_N\limsup_{N\rightarrow\infty}b_N\nonumber\\
\le \limsup_{N\rightarrow\infty}(a_Nb_N)\le\limsup_{N\rightarrow\infty}a_N\limsup_{N\rightarrow\infty}b_N,\nonumber\\
\mathrm{and}~\liminf_{N\rightarrow\infty}a_N+\liminf_{N\rightarrow\infty}b_N\le\liminf_{N\rightarrow\infty}(a_N+b_N)\nonumber\\
\le \liminf_{N\rightarrow\infty}a_N+\limsup_{N\rightarrow\infty}b_N\nonumber\\
\le \limsup_{N\rightarrow\infty}(a_N+b_N)\le\limsup_{N\rightarrow\infty}a_N+\limsup_{N\rightarrow\infty}b_N\nonumber\\
\label{inf_sup}
\end{eqnarray}
\cite{Realanlysiskaczor, RealanlysisRoyden, RealAnalysis_basics_Houshang}. Then using ineq. (\ref{inf_sup}) we obtain,
\begin{eqnarray}
\limsup_{N\rightarrow\infty}\frac{N_{+1}(X_1^{\pi-\theta_1},N_0(X_1,N))}{N}\nonumber\\
\le(\sin^2\frac{\theta_1}{2}+\kappa(X_1^{\pi-\theta_1}=+1,0(X_1)))(\frac{1}{2}+\kappa(X_1=0)).\nonumber
\end{eqnarray}

\section{Case where $\theta_1=\theta_2=0,\pi/2$}\label{theta1eqtheta2eq0app}
For $\theta_1=\theta_2=0$, $N_{+1}(X_1^{\theta_1=0},N_{+1}(X_1,N))=N_{+1}(X_1,N)$, and $N_{+1}(X_1^{\pi-\theta_1=\pi},N_0(X_1,N))=0$.  
\begin{eqnarray}
\Rightarrow F(S(\mathrm{A},M=1)=+1)=\limsup_{N\rightarrow\infty}\frac{N_{+1}(S(\mathrm{A},M=1),N)}{N}\nonumber\\
=\limsup_{N\rightarrow\infty}\frac{N_{+1}(X_1,N)}{N}=1/2+\kappa(X_1=+1).\nonumber
\end{eqnarray}
Alternatively, for $N<\infty$ with $\theta_1=\theta_2=0$, we have $\kappa_N(X_1^{\theta_1=0}=+1,+1(X_1))=0,\kappa_N(X_1^{\pi-\theta_1=\pi}=+1,0(X_1))=0$. Then from ineq. (\ref{FofSM1}) we obtain $F_N(S(\mathrm{A},M=1)=+1)=1/2+\kappa_N(X_1=+1)$. $\Rightarrow \limsup_{N\rightarrow\infty}F_N(S(\mathrm{A},M=1)=+1)=1/2+\kappa(X_1=+1)$.   

For $\theta_1=\theta_2=\pi/2$, 
\begin{eqnarray}
N_{+1}(S(\mathrm{A},M=1),N)=N_{+1}(X_1^{\theta_1=\pi/2},N_{+1}(X_1,N))\nonumber\\
+N_{+1}(X_1^{\pi-\theta_1=\pi/2},N_0(X_1,N))\nonumber\\
=N_{+1}(X_1^{\pi/2},N_{+1}(X_1,N)+N_0(X_1,N))\nonumber\\
=N_{+1}(X_1^{\pi/2},N).\qquad
\label{th1eqth2eqpiby2}
\end{eqnarray}
\begin{eqnarray}
\Rightarrow F(S(\mathrm{A},M=1)=+1)=\limsup_{N\rightarrow\infty}\frac{N_{+1}(S(\mathrm{A},M=1),N)}{N}\nonumber\\
=\limsup_{N\rightarrow\infty}\frac{N_{+1}(X_1^{\pi/2},N)}{N}\nonumber\\
=1/2+\kappa(X_1^{\pi/2}=+1)=1/2+\kappa(X_1=+1).\nonumber
\end{eqnarray}

\section{limit infimum}\label{evalliminfapp}
If we define 
\begin{eqnarray}
F'(S(\mathrm{A},M=1)=+1)=\liminf_{N\rightarrow\infty}\frac{N_{+1}(S(\mathrm{A},M=1),N)}{N}\nonumber\\
\end{eqnarray}
then using ineq. (\ref{inf_sup}), we obtain for the case $\theta_1=\theta_2$ in Eq. (\ref{rak-khas-gan}),
\begin{eqnarray}
&&F'(S(\mathrm{A},M=1)=+1) \ge \frac{1}{2} \nonumber \\
&&+\kappa'(X_1=+1)\left(\cos^2(\theta_1/2)+\kappa'(X_1^{\theta_1}=+1,+1(X_1))\right)\nonumber\\
&& +\kappa'(X_1=0)\left(\sin^2(\theta_1/2)+\kappa'(X_1^{\pi-\theta_1}=+1,0(X_1))\right)\nonumber\\
&&+ \frac{1}{2}\left(\kappa'(X_1^{\theta_1}=+1,+1(X_1))+\kappa'(X_1^{\pi-\theta_1}=+1,0(X_1))\right)\nonumber
\end{eqnarray}
where $\kappa'(\cdots)$'s correspond to limit infimum.

\section{On the observability of content dependent fluctuation}\label{observbltyofcontentdepflucapp}
The fact that limiting relative frequency cannot always converge pointwise to a given constant value (real number) proves the existence of fluctuation term $\kappa(.)$ and hence content dependent fluctuation (i.e., fluctuation of $F(S(\mathrm{A},M=1)=+1)$ depends on $\theta_1,\theta_2$) (see Appendix \ref{kappadependsonthetaapp}). However it should be noted that observing content dependent fluctuation may not be as difficult as observing violation of the requirement for pointwise convergence e.g., observing the violation of $|N_{+1}(X_1,N)/N-1/2|<\epsilon$ where $\epsilon>0,N<\infty$, usually becomes difficult if we choose $N$ sufficiently large (this is justified by the stabilization of relative frequency which is an experimental fact \cite{Prob_book_AlanGut}). This is because, former requires observing relative fluctuation only, which do not depend only on rare events unlike the latter which depends only on rare events.

\section{Perfect anti-correlation of singlet in FQM}\label{perfectanticorsingltapp}
We can rewrite $\lim_{N\rightarrow\infty}S(\sigma_{z}^\mathcal{A}\sigma_{z}^\mathcal{B},N)$, defined in Sec. \ref{Furtheraspects} as follows:
\begin{eqnarray}
\lim_{N\rightarrow\infty}S(\sigma_{z}^\mathcal{A}\sigma_{z}^\mathcal{B},N)=\lim_{N\rightarrow\infty}\nonumber\\
\frac{N_{+1}(S(\sigma_{z}^\mathcal{A}\sigma_{z}^\mathcal{B},M=1),N)-N_{-1}(S(\sigma_{z}^\mathcal{A}\sigma_{z}^\mathcal{B},M=1),N)}{N}.\nonumber
\end{eqnarray}
We have
\begin{eqnarray}
N_{+1}(S(\sigma_{z}^\mathcal{A}\sigma_{z}^\mathcal{B},M=1),N)\nonumber\\
=N_{+1+1}(S(\sigma_{z}^\mathcal{A}\sigma_{z}^\mathcal{B},M=1),N)\nonumber\\
+N_{-1-1}(S(\sigma_{z}^\mathcal{A}\sigma_{z}^\mathcal{B},M=1),N)=0+0.\nonumber\\
N_{-1}(S(\sigma_{z}^\mathcal{A}\sigma_{z}^\mathcal{B},M=1),N)\nonumber\\
=N_{+1-1}(S(\sigma_{z}^\mathcal{A}\sigma_{z}^\mathcal{B},M=1),N)\nonumber\\
+N_{-1+1}(S(\sigma_{z}^\mathcal{A}\sigma_{z}^\mathcal{B},M=1),N)\nonumber\\
=N_{+1}(\sigma_{z}^\mathcal{A},N)+N_{-1}(\sigma_{z}^\mathcal{A},N)=N.\nonumber
\end{eqnarray}
\begin{eqnarray}
\Rightarrow\lim_{N\rightarrow\infty} S(\sigma_{z}^\mathcal{A}\sigma_{z}^\mathcal{B},N)\nonumber\\
=-\lim_{N\rightarrow\infty}\frac{N_{+1}(\sigma_{z}^\mathcal{A},N)+N_{-1}(\sigma_{z}^\mathcal{A},N)}{N}=-1.\nonumber
\end{eqnarray}

\section{The case when $1\ll N<\infty$}\label{finiteNcaseapp}
Consider the case when $1\ll N<\infty$. Define
 \begin{eqnarray}
F_N(X=+1)=\frac{N_{+1}(X,N)}{N}:=  \frac{1}{2}+\kappa_N(X=+1)
\end{eqnarray}
where $\kappa_N(X=+1)$ is a random variable which takes 
values in 
\([-\epsilon_N,\delta_N]
~(\epsilon_N>0,~\delta_N>0)$. 
Then the expression corresponding to ineq. (\ref{N1byN}) will be the following, 
\begin{eqnarray}
\frac{N_{+1}(X_1^{\theta_1},N_{+1}(X_1,N))}{N_{+1}(X_1,N)}\frac{N_{+1}(X_1,N)}{N}\nonumber\\
=(\cos^2(\theta_1/2)+\kappa_N(X_1^{\theta_1}=+1,+1(X_1)))\nonumber\\
\times(1/2+\kappa_N(X_1=+1)),~~~~
\label{N1byN_0}
\end{eqnarray}
for $N_{+1}(X_1,N)>0$. This shows that, in all the results derived in the main text, we just have to replace $\kappa(\cdots)$'s with the corresponding $\kappa_N(\cdots)$'s, and inequalities become equalities. Of course the constraint that the terms in the denominators should be greater than zero should be satisfied (like $N_{+1}(X_1,N)>0$ in Eq. (\ref{N1byN_0})). Further note that $\limsup_{N\rightarrow\infty}F_N(X=+1)=F(X=+1)$ and hence $\limsup_{N\rightarrow\infty}\kappa_N(X=+1)=\kappa(X=+1)$ as required. Similarly we obtain $\limsup_{N\rightarrow\infty}\kappa_N(\cdots)=\kappa(\cdots)$. \\Further note that when $N$ is very small (say e.g., $1\le N\le 10$), then both $F_N(S(\mathrm{A},M=1)=+1)$ and $F_N(S(\mathrm{B},M=1)=+1)$ will easily saturate i.e., will easily take maximum and minimum possible values which are 1 and 0 respectively. Hence Bob cannot distinguish. Fig. \ref{plotfluctn} is helpful in understanding this point.

\section{Case where $\theta_2\ne \theta_1$ in Eq. (\ref{rak-khas-gan}) }\label{theta1neqtheta2caseapp}
Let $M=1$.
Let $N_{x_1x_1^\Theta}((X_1,X^\Theta_1),N)$ be the number of $X_1=x_1$ and $X^\Theta_1=x_1^\Theta$ outcomes in $N$ independent trials each of $X_1$ and $X^\Theta_1$. Then we have the following identity
$N_{x_1x_1^\Theta}((X_1,X_1^\Theta),N)=N_{x_1}(X_1,N_{x_1^\Theta}(X_1^\Theta,N))$ ($\because$ events are independent) where 
$N_{x_1^\Theta}(X_1^\Theta,N)$ is the number of $x_1^\Theta$ outcomes in $N$ independent trials of $X_1^\Theta$, 
$x_1=+1,0;x_1^\Theta=\theta_1,\theta_2$; and $N_{\theta_1}(X_1^\Theta,N)+N_{\theta_2}(X_1^\Theta,N)=N$. Further we have
\begin{eqnarray}
N_{+1}(S(\mathrm{A},M=1),N)=N_{+1}(X_1^{\theta_1},N_{+1\theta_1}((X_1,X^\Theta_1),N))\nonumber\\
+N_{+1}(X_1^{\theta_2},N_{+1\theta_2}((X_1,X^\Theta_1),N))\nonumber\\
+N_{+1}(X_1^{\pi-\theta_1},N_{0\theta_1}((X_1,X^\Theta_1),N))\nonumber\\
+N_{+1}(X_1^{\pi-\theta_2},N_{0\theta_2}((X_1,X^\Theta_1),N))\nonumber
\end{eqnarray}
where $N_{+1\theta_1}((X_1,X^\Theta_1),N)+N_{+1\theta_2}((X_1,X^\Theta_1),N)+N_{0\theta_1}((X_1,X^\Theta_1),N)+N_{0\theta_2}((X_1,X^\Theta_1),N)=N$. Then using ineq. (\ref{inf_sup}) we obtain
\begin{eqnarray}
\limsup_{N\rightarrow\infty}\frac{N_{+1}(X_1^{\theta_1},N_{+1\theta_1}((X_1,X^\Theta_1),N))}{N}\nonumber\\
=\limsup_{N\rightarrow\infty}\frac{N_{+1}(X_1^{\theta_1},N_{+1\theta_1}((X_1,X^\Theta_1),N))}{N_{+1\theta_1}((X_1,X^\Theta_1),N)}\nonumber\\
\times\frac{N_{+1\theta_1}((X_1,X^\Theta_1),N)}{N}\nonumber\\
=\limsup_{N\rightarrow\infty}\frac{N_{+1}(X_1^{\theta_1},N_{+1}(X_1,N_{\theta_1}(X^\Theta_1,N)))}{N_{+1}(X_1,N_{\theta_1}(X^\Theta_1,N))}\nonumber\\
\times\frac{N_{+1}(X_1,N_{\theta_1}(X^\Theta_1,N))}{N_{\theta_1}(X^\Theta_1,N)}\frac{N_{\theta_1}(X^\Theta_1,N)}{N}\nonumber\\
\le\limsup_{N\rightarrow\infty}\frac{N_{+1}(X_1^{\theta_1},N_{+1}(X_1,N_{\theta_1}(X^\Theta_1,N)))}{N_{+1}(X_1,N_{\theta_1}(X^\Theta_1,N))}\nonumber\\
\times\limsup_{N\rightarrow\infty}\frac{N_{+1}(X_1,N_{\theta_1}(X^\Theta_1,N))}{N_{\theta_1}(X^\Theta_1,N)}\limsup_{N\rightarrow\infty}\frac{N_{\theta_1}(X^\Theta_1,N)}{N},\nonumber
\end{eqnarray}
for $N_{+1}(X_1,N_{\theta_1}(X^\Theta_1,N\rightarrow\infty))>0,N_{\theta_1}(X^\Theta_1,N\rightarrow\infty)>0$. Substituting $\theta_1=0,\theta_2=\pi$ in the above expression, we obtain
\begin{eqnarray}
\limsup_{N\rightarrow\infty}\frac{N_{+1}(X_1^{\theta_1},N_{+1\theta_1}((X_1,X^\Theta_1),N))}{N}\nonumber\\
\le (1/2+\kappa(X_1=+1,\theta_1(X_1^\Theta)))(1/2+\kappa(X^\Theta_1=\theta_1))\nonumber
\end{eqnarray}
($\because N_{+1}(X_1^{\theta_1=0},N_{+1}(X_1,N_{\theta_1}(X^\Theta_1,N)))=N_{+1}(X_1,N_{\theta_1}(X^\Theta_1,N))$).
Similarly we obtain 
\begin{eqnarray}
\limsup_{N\rightarrow\infty}\frac{N_{+1}(X_1^{\theta_2},N_{+1\theta_2}((X_1,X^\Theta_1),N))}{N}\nonumber\\
\le\limsup_{N\rightarrow\infty}\frac{N_{+1}(X_1^{\theta_2},N_{+1}(X_1,N_{\theta_2}(X^\Theta_1,N)))}{N_{+1}(X_1,N_{\theta_2}(X^\Theta_1,N))}\nonumber\\
\times\limsup_{N\rightarrow\infty}\frac{N_{+1}(X_1,N_{\theta_2}(X^\Theta_1,N))}{N_{\theta_2}(X^\Theta_1,N)}\limsup_{N\rightarrow\infty}\frac{N_{\theta_2}(X^\Theta_1,N)}{N}.\nonumber
\end{eqnarray}
Substituting $\theta_1=0,\theta_2=\pi$ in the above expression, we obtain
\begin{eqnarray}
\limsup_{N\rightarrow\infty}\frac{N_{+1}(X_1^{\theta_2},N_{+1\theta_2}((X_1,X^\Theta_1),N))}{N}=0\nonumber
\end{eqnarray}
($\because N_{+1}(X_1^{\theta_2=\pi},N_{+1}(X_1,N_{\theta_2}(X^\Theta_1,N)))=0$). Similarly
\begin{eqnarray}
\limsup_{N\rightarrow\infty}\frac{N_{+1}(X_1^{\pi-\theta_1},N_{0\theta_1}((X_1,X^\Theta_1),N))}{N}\nonumber\\
\le\limsup_{N\rightarrow\infty}\frac{N_{+1}(X_1^{\pi-\theta_1},N_{0}(X_1,N_{\theta_1}(X^\Theta_1,N)))}{N_{0}(X_1,N_{\theta_1}(X^\Theta_1,N))}\nonumber\\
\times\limsup_{N\rightarrow\infty}\frac{N_{0}(X_1,N_{\theta_1}(X^\Theta_1,N))}{N_{\theta_1}(X^\Theta_1,N)}\limsup_{N\rightarrow\infty}\frac{N_{\theta_1}(X^\Theta_1,N)}{N}.\nonumber
\end{eqnarray}
Substituting $\theta_1=0,\theta_2=\pi$ in the above expression, we obtain
\begin{eqnarray}
\limsup_{N\rightarrow\infty}\frac{N_{+1}(X_1^{\pi-\theta_1},N_{0\theta_1}((X_1,X^\Theta_1),N))}{N}=0\nonumber
\end{eqnarray}
($\because N_{+1}(X_1^{\pi-\theta_1=\pi},N_{0}(X_1,N_{\theta_1}(X^\Theta_1,N)))=0$). Similarly
\begin{eqnarray}
\limsup_{N\rightarrow\infty}\frac{N_{+1}(X_1^{\pi-\theta_2},N_{0\theta_2}((X_1,X^\Theta_1),N))}{N}\nonumber\\
\le\limsup_{N\rightarrow\infty}\frac{N_{+1}(X_1^{\pi-\theta_2},N_{0}(X_1,N_{\theta_2}(X^\Theta_1,N)))}{N_{0}(X_1,N_{\theta_2}(X^\Theta_1,N))}\nonumber\\
\times\limsup_{N\rightarrow\infty}\frac{N_{0}(X_1,N_{\theta_2}(X^\Theta_1,N))}{N_{\theta_2}(X^\Theta_1,N)}\limsup_{N\rightarrow\infty}\frac{N_{\theta_2}(X^\Theta_1,N)}{N}.\nonumber
\end{eqnarray}
Substituting $\theta_1=0,\theta_2=\pi$ in the above expression, we obtain
\begin{eqnarray}
\limsup_{N\rightarrow\infty}\frac{N_{+1}(X_1^{\pi-\theta_2},N_{0\theta_2}((X_1,X^\Theta_1),N))}{N}\nonumber\\
\le(1/2+\kappa(X_1=0,\theta_2(X_1^\Theta)))(1/2+\kappa(X^\Theta_1=\theta_2))\nonumber
\end{eqnarray}
($\because N_{+1}(X_1^{\pi-\theta_2=0},N_{0}(X_1,N_{\theta_2}(X^\Theta_1,N)))=N_{0}(X_1,N_{\theta_2}(X^\Theta_1,N))$). Substituting the above expressions into Eq. (\ref{FofSM1gen}), we obtain for the case $\theta_1=0,\theta_2=\pi$, the following expression 
\begin{eqnarray}
&&F(S(\mathrm{A},M=1)=+1)\le \frac{1}{2}+\frac{\kappa(X_1^\Theta=\theta_1)+\kappa(X_1^\Theta=\theta_2)}{2}\nonumber\\
&&+\big(\kappa(X_1=+1,\theta_1(X_1^\Theta))+\kappa(X_1=0,\theta_2(X_1^\Theta))\big)/2\nonumber\\
&&+\kappa(X_1^\Theta=\theta_1)\kappa(X_1=+1,\theta_1(X_1^\Theta))\nonumber\\
&&+\kappa(X_1^\Theta=\theta_2)\kappa(X_1=0,\theta_2(X_1^\Theta)).
\label{FofSAmain_0}
\end{eqnarray}

Further, we can rewrite, 
	\begin{eqnarray}
\limsup_{N\rightarrow\infty}S(\mathrm{A},N)=\limsup_{N\rightarrow\infty}\nonumber\\
	\frac{N_{+1}(S(\mathrm{A},M=1),N)-(N-N_{+1}(S(\mathrm{A},M=1),N))}{N}\nonumber\\
	=2F(S(\mathrm{A},M=1)=+1)-1\nonumber\\
=	\kappa(X_1^\Theta=\theta_1)+\kappa(X_1^\Theta=\theta_2)\nonumber\\
	+\kappa(X_1=+1,\theta_1(X_1^\Theta))+\kappa(X_1=0,\theta_2(X_1^\Theta))\nonumber\\
	+2\kappa(X_1^\Theta=\theta_1)\kappa(X_1=+1,\theta_1(X_1^\Theta))\nonumber\\
	+2\kappa(X_1^\Theta=\theta_2)\kappa(X_1=0,\theta_2(X_1^\Theta))\qquad
	\label{SofA}
\end{eqnarray}
where we used Eq. (\ref{FofSM1gen}) and expression (\ref{FofSAmain_0}). Similarly we can rewrite, 
\begin{eqnarray}
\limsup_{N\rightarrow\infty}S(\mathrm{B},N)=\limsup_{N\rightarrow\infty}\nonumber\\
	\frac{N_{+1}(S(\mathrm{B},M=1),N)-(N-N_{+1}(S(\mathrm{B},M=1),N))}{N}\nonumber\\
	=2\kappa(X_1=+1)=2\kappa(X_1^\Theta=\theta_1)~~~~~~~~~~
	\label{SBNinfty}
	\end{eqnarray}
 where we used Eq. (\ref{PofS_B}).

 \section{Associating normal distribution with the fluctuation of $\kappa_N(...)$ terms for practical purposes}\label{normaldistrbforfappapp}
 Here we quantify (for practical purposes) using KQM, the content dependent fluctuation in $S(\mathrm{A},N)$, and the fluctuation of $S(\mathrm{B},N)$. 

\subsection{$S(\mathrm{A},N) \approx\kappa_{M_1}(X_1=+1)+\kappa_{M_2}(X_1=0)+ 4\kappa_N(X_1^\Theta=\theta_1)\kappa_{N}(X_1=+1)$ for $N\gg1,\theta_1=0,\theta_2=\pi$}
In the case $\theta_1=0,\theta_2=\pi$ in Eq. (\ref{rak-khas-gan}), for $N\gg 1$, we can make following approximations:
\begin{eqnarray}
\frac{N_{+1}(X_1,N_{\theta_1}(X_1^\Theta,N))}{N_{\theta_1}(X_1^\Theta,N)}=\frac{1}{2}+\kappa_N(X_1=+1,\theta_1(X_1^\Theta))\nonumber\\
\approx\frac{N_{+1}(X_1,M_1)}{M_1}=\frac{1}{2}+\kappa_{M_1}(X_1=+1),\qquad\label{M1}\\
\frac{N_{0}(X_1,N_{\theta_2}(X_1^\Theta,N))}{N_{\theta_2}(X_1^\Theta,N)}=\frac{1}{2}+\kappa_N(X_1=0,\theta_2(X_1^\Theta))\nonumber\\
\approx\frac{N_{0}(X_1,M_2)}{M_2}=\frac{1}{2}+\kappa_{M_2}(X_1=0)\qquad\label{M2}
\end{eqnarray}
where $M_1=N/2,M_2=N/2$ (for convenience we have assumed $N$ to be even). It is important to note that $M_1$ number of trials of $X_1$ are independent and different from $M_2$ number of trials of $X_1$. This is represented by denoting each of the $N/2$ number of trials in Eqs. (\ref{M1}, \ref{M2}) using different symbols i.e., $M_1$ and $M_2$. Further 
\begin{eqnarray}
\frac{N_{+1}(X_1,M_1)}{M_1}-\frac{N_{0}(X_1,M_2)}{M_2}=2\frac{N_{+1}(X_1,N)}{N}-1\nonumber\\
\Rightarrow\kappa_{M_1}(X_1=+1)-\kappa_{M_2}(X_1=0)=2\kappa_N(X_1=+1).\nonumber
\end{eqnarray}
 And $\kappa_N(X_1^\Theta=\theta_{1})=-\kappa_N(X_1^\Theta=\theta_2)$.
Substituting these into the finite $N$ expression corresponding to expression (\ref{SofA}), we obtain
 	\begin{eqnarray}
  S(\mathrm{A},N) \approx\kappa_{M_1}(X_1=+1)+\kappa_{M_2}(X_1=0)\nonumber\\
  +4\kappa_N(X_1^\Theta=\theta_1)\kappa_{N}(X_1=+1).
 	\label{S_procdrA}
 	\end{eqnarray}

\subsection{Plotting the density of $F_N(S(A/B,M=1)=+1)$}
To experimentally study the fluctuation of $F_N(S(A/B,M=1)=+1)$, 
we should repeat the experiment $n$ times and plot the density of $F_N(S(A/B,M=1)=+1)$ versus $F_N(S(A/B,M=1)=+1)$ where density of $F_N(S(A/B,M=1)=+1)$ is nothing but the ratio of number of times we get $F_N(S(A/B,M=1)=+1)=y$ in $n$ repetitions and $(n\times \delta F_N(S(A/B,M=1)=+1))$ where $y\in[0,1]$ and $\delta F_N(S(A/B,M=1)=+1)(=1/N)$ is the step size. For example, consider the simplest case of plotting the density of 
\begin{eqnarray}
F_N(S(B,M=1)=+1)\nonumber\\
=N_{+1}(S(B,M=1),N)/N=N_{+1}(X^{\pi/2}_1,N)/N\nonumber\\
=1/2+\kappa_N(X^{\pi/2}_1=+1).
\end{eqnarray}
$N_{+1}(X^{\pi/2}_1,N)$ takes value $y'\in\{0,1,2,...,N\}$ and hence $F_N(S(B,M=1)=+1)$ takes value $y\in\{0,1/N,2/N,...,1\}$. Hence $F_N(S(B,M=1)=+1)$ tends to become a continuous random variable in the limit $N\rightarrow\infty$. Now we repeat $n$ times the experiment involving $N$ trials of $X^{\pi/2}_1$. Then we calculate the ratio of number of times we get $F_N(S(B,M=1)=+1)=y$ in $n$ repetitions and $(n\times(1/N))$. Then we plot this ratio versus $y$. For $N\gg 1$, we will obtain this plot to be approximately a Gaussian centered around $1/2$ (this we know from actual experiment) (KQM predicts that Gaussian will have mean $1/2$ and variance $1/(4N)$). This is how we can experimentally study the fluctuation of $F_N(S(B,M=1)=+1)$, and hence the fluctuation of $\kappa_N(X^{\pi/2}_1=+1)$. Similarly, we can experimentally study the fluctuation of $F_N(S(A,M=1)=+1)$. FQM predicts that the fluctuation of $F_N(S(A,M=1)=+1)$ will be different from that of $F_N(S(B,M=1)=+1)$.

 Now we can safely (i.e., without loss of any fundamental content-dependent fluctuations) bring in KQM for practical purposes and quantify the fluctuation of $\kappa_N(...)$ terms as follows. It is an experimental fact that if we plot the density of $F_N(X=+1)$ versus $F_N(X=+1)$, we obtain approximately a Gaussian function centered approximately around 1/2. Hence it is reasonable for 
 practical purposes to associate a normal probability density function with the fluctuation of $\kappa_N(...)$ terms, i.e.,
\begin{eqnarray}
f(\kappa_N(X_1=x_1))\approx\frac{1}{\sqrt{2\pi\mathrm{Var}(X_1)/N}}\exp(\frac{-\kappa_N(X_1=x_1)^2}{2\mathrm{Var}(X_1)/N}),\nonumber\\
\label{kapp_distrbtn_normalapprox}
\end{eqnarray} 
where $f(Z)$ is the probability density function of the random variable $Z$, and $\mathrm{Var}(X_1)=\expec{X_1^2}-\expec{X_1}^2=1/4$ is the variance of $X_1$. Note that we can associate mean zero with every $\kappa(...)$ term. This is because, according to KQM, 
\begin{eqnarray}
\expec{F_N(X=+1)}=\expec{X}=1/2\Rightarrow\expec{\kappa_N(X=+1)}=0,\nonumber\\
\mathrm{Var}(F_N(X=+1))=\frac{\mathrm{Var}(X)}{N}=\frac{1}{4N}=\mathrm{Var}(\kappa_N(X=+1)),\nonumber\\
\expec{F_N(X=+1)}=\expec{F_N(X^\Theta=\theta_1)},\nonumber\\
\mathrm{Var}(F_N(X=+1))=\mathrm{Var}(F_N(X^\Theta=\theta_1)).\qquad
\label{FNinKQM}
\end{eqnarray}
Further, it is important to note that from a foundational perspective, fluctuation of $\kappa_N(X_1=+1)$ do not vanish even in the limit $N\rightarrow\infty$, contrary to the approximation in (\ref{kapp_distrbtn_normalapprox}), which becomes a ``delta function''. This is due to no pointwise convergence of LRF, always to 1/2. Note that for notational convenience, we are using the same symbol for the random variables $\kappa_N(...)$'s and also the values they take. Its meaning should be understood from the context of usage. To associate an approximate probability density function with $4\kappa_N(X_1=+1)\kappa_N(X_1^\Theta=\theta_1)$ in expression (\ref{S_procdrA}), we proceed as follows:
\begin{eqnarray}
f(\zeta=4\kappa_N(X_1=+1)\kappa_N(X_1^\Theta=\theta_1))\nonumber\\
\approx\int_{-\infty}^{\infty}d\kappa_N(X_1^\Theta=\theta_1)~f(\zeta|\kappa_N(X_1^\Theta=\theta_1))f(\kappa_N(X_1^\Theta=\theta_1)),\nonumber\\
\label{kappXXtheta_distrbtn_normalapprox}
\end{eqnarray}
where $f(\zeta,\kappa_N(X_1^\Theta=\theta_1))=f(\zeta|\kappa_N(X_1^\Theta=\theta_1))f(\kappa_N(X_1^\Theta=\theta_1))$. Note that $\zeta$ depends on $\kappa_N(X_1^\Theta=\theta_1)$ and hence $f(\zeta|\kappa_N(X_1^\Theta=\theta_1))\ne f(\zeta)$.
\\\textbf{Theorem-1} \cite{Sheldon_Ross_probability_book}: If $X$ is a normally distributed random variable with mean $\mu$ and variance $\sigma^2$, then $Y=aX+b$ is also a normally distributed random variable with mean $a\mu+b$ and variance $a^2\sigma^2$ where $a,b$ are constants.\\Using approximations (\ref{kapp_distrbtn_normalapprox}) and (\ref{kappXXtheta_distrbtn_normalapprox}), and theorem-1, we obtain
\begin{widetext}
\begin{eqnarray}
f(\zeta=4\kappa_N(X_1=+1)\kappa_N(X_1^\Theta=\theta_1))\nonumber\\
\approx\int_{-\infty}^{\infty}\frac{d\kappa_N(X_1^\Theta=\theta_1)}{\sqrt{2\pi\mathrm{Var}(X_1)/N}}\exp(\frac{-\kappa_N(X_1^\Theta=\theta_1)^2}{2\mathrm{Var}(X_1)/N})\sqrt{\frac{N}{32\pi\kappa_N(X_1^\Theta=\theta_1)^2\mathrm{Var}(X_1)}}\exp(\frac{-N\zeta^2}{32\kappa_N(X_1^\Theta=\theta_1)^2\mathrm{Var}(X_1)}),
\label{kappXXtheta_distrbtn_normalapprox1}
\end{eqnarray} 
\end{widetext}
where $\mathrm{Var}(X_1)=1/4$, and where we have used the fact that $\kappa_N(X_1=+1)$ and $\kappa_N(X_1^\Theta=\theta_1)$ are independent random variables and that the same variance ($=1/(4N)$) must be associated with each of them (because $X_1$ and $X_1^\Theta$ differ only in the value assigned to their outcomes. See Eqs. (\ref{FNinKQM}) in this regard). If $\kappa_N(X_1=+1)$ and $\kappa_N(X_1^\Theta=\theta_1)$ were not independent, then for a given value of $\kappa_N(X_1^\Theta=\theta_1)$, the probability distribution which we can associate with $\kappa_N(X_1=+1)$ will depend on the given value of $\kappa_N(X_1^\Theta=\theta_1)$ as well. There is no analytical solution to the integral (\ref{kappXXtheta_distrbtn_normalapprox1}) (see \cite{Distrbtn_XY_each_Normal} in this regard, and for further details regarding approximate and numerical solutions to the integral), and in particular the distribution is not normal. Further $\eta=\kappa_{M_1}(X_1=+1)+\kappa_{M_2}(X_1=0)$ is normally distributed with mean $0$ and variance $1/N$ \cite{Sheldon_Ross_probability_book}. And $\eta,\zeta$ are not independent. Hence $S(\mathrm{A},N)$ cannot be normally distributed. We also have $S(\mathrm{B},N)=2\kappa_N(X_1=+1)$ (Eq. (\ref{SBNinfty})). But 
\begin{eqnarray}
f(2\kappa_N(X_1=+1))\nonumber\\
\approx\frac{1}{\sqrt{8\pi\mathrm{Var}(X_1)/N}}\exp(\frac{-\kappa_N(X_1=+1)^2}{8\mathrm{Var}(X_1)/N}).~~\quad
\label{kapp_distrbtn_normalapproxProcdrB}
\end{eqnarray} 
Hence the fluctuations of sample means around 0 are different in the two preparation procedures A and B.

\bibliographystyle{apsrev4-1}

\begin{thebibliography}{24}


\bibitem{Prob_measr_book} P. Billingsley, \emph{Probability and Measure} (Wiley, 1995).

\bibitem{Prob_book_AlanGut} A. Gut, \emph{Probability: A Graduate Course} (Springer, 2005).

\bibitem{Prob_book_Chow} Y. S. Chow and H. Teicher, \emph{Probability theory: Independence, interchangeability, martingales} (Springer,
2005).

\bibitem{Sheldon_Ross_probability_book} S. Ross, \emph{A first course in probability} (Pearson, 
2010).

\bibitem{Prob_book_weigh_odd} D. Williams, \emph{Weighing the Odds: A Course in Probability and Statistics} (Cambridge, 2010).

\bibitem{QM_Griffiths} D. J. Griffiths, \emph{Introduction to Quantum Mechanics} (Prentice Hall, 1995).

\bibitem{Peres_QM_book} A. Peres, \emph{Quantum Theory: Concepts and Methods}
(Kluwer Academic, 2002).

\bibitem{QM_Cohen} C. Cohen-Tannoudji, B. Diu, and F. Lalo{\"e}, \emph{Quantum Mechanics}, Vol. I
(Wiley, 2005).

\bibitem{QM_book_RShankar}R. Shankar, \emph{Principles of Quantum Mechanics} (Springer, 2008).

\bibitem{Mathews-Venkatesan} P. M. Mathews and K. Venkatesan, \emph{A Textbook of Quantum Mechanics} (Tata McGraw Hill Education, 2010).


\bibitem{circularLLNSpanos2013} A. Spanos, Synthese \textbf{190}, 1555 (2013).

\bibitem{Emperor_new_mind_penrose}	R. Penrose, \emph{The Emperor's New Mind} (Oxford university press, 2006).

\bibitem{Goedel-incompleteness}P. Raatikainen, \emph{The Stanford Encyclopedia of Philosophy: G{\"o}del's Incompleteness Theorems} (Stanford University,
2018).

\bibitem{vonMises_work_review} H. Cramer, The Annals of Mathematical Statistics \textbf{24}, 657 (1953).

\bibitem{vonMises_prob_book}R. Von Mises, \emph{Probability, Statistics, and Truth} (Dover,
1981).

\bibitem{probInterprtn} A. H{\'a}jek,  \emph{The Stanford Encyclopedia of Philosophy: Interpretations of Probability} (Stanford University, 2019).


\bibitem{pathwiseKARANDIKAR199511} R. L. Karandikar, Stochastic Processes and their Applications \textbf{57}, 11 (1995).

\bibitem{FollmerBook}D. Sondermann, \emph{Introduction to Stochastic Calculus for Finance} (Springer,
2006).

\bibitem{Hobson2011}D. Hobson, \emph{Paris-Princeton Lectures on Mathematical Finance 2010} (Springer, 2011, pp. 267-318).

\bibitem{pathwiseFinanceCandia} C. Riga, arXiv:1602.04946v1[q-fin.MF] (2016).

\bibitem{pathwiseRamaConnt} A. Ananova and R. Cont, Journal de Math{\'e}matiques Pures et Appliqu{\'e}es \textbf{107}, 737 (2017).

\bibitem{Bellvio4Popescu1994} S. Popescu and D. Rohrlich, Foundations of Physics \textbf{24},
379 (1994).

\bibitem{No_communication_Peres} A. Peres and D. R. Terno, Rev. Mod. Phys. \textbf{76}, 93 (2004).

\bibitem{No_signal_in_time_Kofler} J. Kofler and {\v C}. Brukner, Phys. Rev. A \textbf{87}, 052115
(2013).

\bibitem{banshi-shune-aar-kaj-nai} N. D. Mermin, Rev. Mod. Phys. \textbf{65}, 803 (1993).

\bibitem{Finetti_imprcProb} P. Vicig and T. Seidenfeld, International Journal of Approximate Reasoning \textbf{53}, 1115 (2012).

\bibitem{OckhamRazorStanEncyclo} P. V. Spade and C. Panaccio, \emph{The Stanford Encyclopedia of Philosophy: William of Ockham} (Stanford University, 2019).


\bibitem{RealanlysisRoyden} H. L. Royden, \emph{Real Analysis} (The Macmillan company, second edition, 1968).

\bibitem{Math_analysis_book} W. Rudin, \emph{Principles of mathematical analysis} (McGraw Hill, 1976).

\bibitem{RealAnalysis_Apostl} T. M. Apostol, \emph{Mathematical Analysis} (Narosa,
1985).

\bibitem{RealAnalysis_basics_Houshang} H. H. Sohrab, \emph{Basic real analysis} (Springer, 2006).

\bibitem{Topology_Munkres} J. R. Munkres, \emph{Topology} 
(Prentice Hall, 
2007).


\bibitem{Amritava-Gupta} A. Gupta, \emph{Introduction To Mathematical Analysis} 
(Academic Publishers, 
2016).

\bibitem{KnightsBook} F. H. Knight, \emph{Risk, Uncertainty, and Profit} (Houghton Mifflin Company, 1921).

\bibitem{KnightUncertntyFinancePaper} Y. Ben-Haim  and M. Demertzis, Economics: The Open-Access, Open-Assessment E-Journal \textbf{10(2016-23)}, 1 (2016).





\bibitem{No_pointwise_convrg}
If \(\lim_{N \to \infty} N_{+1} (X, N )/N = 1/2\), then
it implies 
that
\(\forall 0 < \epsilon
< 1/2, \exists M < \infty : \forall N >
M, |N_{+1} (X, N )/N - 1/2| < \epsilon\). However, as \(N <
\infty\), all possible outcomes will be realised with nonzero
(positive) chance or possibility or likeliness, upon repeating
the experiment many times. Hence the requirement for
pointwise convergence to 1/2 is not always satisfied (e.g., for
\(N_{+1} (X, N ) = N\)).




\bibitem{ensembl_discrm_NMR_Found_phy} G. L. Long, Y.-F. Zhou, J.-Q. Jin, Y. Sun, and H.-
W. Lee, quant ph/0408079v3; Foundations of Physics \textbf{36},
1217 (2006).

\bibitem{ensembl_discrm_dEspanag_PLA} J. Tolar and P. H{\'a}j{\'i}{\v c}ek, Phys. Lett. A \textbf{353}, 19 (2006).


\bibitem{Popescu_conjctr} S. Popescu, arXiv:1811.05472v1 [quant-ph] (2018).


\bibitem{AOC_lindep} In the limit $\mathcal{M}\rightarrow\infty$, due to Anderson's orthogonality catastrophe (AOC) \cite{ortho_catstrpAnderson} we obtain $|\inpr{\psi_j^\mathrm{A}}{\psi_k^\mathrm{B}}|\rightarrow 0~\forall j,k$. However the states $\ket{\psi_k^\mathrm{B}}$'s will be still linearly dependent on $\ket{\psi_j^\mathrm{A}}$'s. This is a seemingly strange property exhibited only by infinite tensor product non-separable Hilbert spaces \cite{von_Neumann_colectdwork_infinitedirectprodct}. And hence within KQM, in spite of AOC, it is still not possible to distinguish between the two preparation procedures A and B \cite{USD_chefls_linind}. Proof of this will be published elsewhere.


\bibitem{ortho_catstrpAnderson} P. W. Anderson, Phys. Rev. Lett. \textbf{18}, 1049 (1967).

\bibitem{von_Neumann_colectdwork_infinitedirectprodct} J. von Neumann, \emph{John von Neumann collected works Vol. III, chapter 6 ``On infinite direct products''} (Pergamon press, 1976).

\bibitem{USD_chefls_linind} A. Chefles, Phys. Lett. A \textbf{239}, 339 (1998).

\bibitem{superactvtn_peres} A. Peres, Phys. Rev. A \textbf{54}, 2685 (1996).

\bibitem{superactvtn_Sen} A. Sen(De), U. Sen, {\v C}. Brukner, V. Bu{\v z}ek, and M. {\.Z}ukowski, Phys. Rev. A \textbf{72}, 042310 (2005).

\bibitem{superactivtn_palazuola} C. Palazuelos, Phys. Rev. Lett. \textbf{109}, 190401 (2012).


\bibitem{Realanlysiskaczor} W. J. Kaczor and M. T. Nowak, \emph{Problems in Mathematical Analysis I} (American Mathematical Society,
2009).



\bibitem{Teleport_benet_orginal} C. H. Bennett, G. Brassard, C. Cr\'epeau, R. Jozsa, A. Peres, and W. K. Wootters,  Phys. Rev. Lett. \textbf{70}, 1895 (1993).

\bibitem{quant_info_neilson_chuang} M. A. Nielsen and I. L. Chuang, \emph{Quantum Computation and Quantum Information}, (Cambridge University Press, 2010).

\bibitem{noclone_buzek5by6} V. Bu{\v z}ek  and M. Hillery, Phys. Rev. A \textbf{54}, 1844 (1996).

\bibitem{BB84_original_paper} C. H. Bennett and G. Brassard, Proceedings of International conference on computers, systems and signal processing Bangalore, India \textbf{1}, 175 (1984).


\bibitem{CTC_Brunn_breakbb84} T. A. Brun, J. Harrington, and M. M. Wilde,
Phys. Rev. Lett. \textbf{102}, 210402 (2009).

\bibitem{CTC_Benett}C. H. Bennett, D. Leung, G. Smith, and J. A. Smolin, Phys. Rev. Lett. \textbf{103}, 170502 (2009).

\bibitem{QKDmongmyPawlosky} M. Paw\l{}owski, Phys. Rev. A \textbf{82}, 032313 (2010).

\bibitem{Popescunonlocalbeyndqmreviewnpj} S. Popescu, Nat. Phys. \textbf{10}, 264 (2014).

\bibitem{PTsym_vio_nosignal} Y.-C. Lee, M.-H. Hsieh, S. T. Flammia, and R.-K. Lee,
Phys. Rev. Lett. \textbf{112}, 130404 (2014).

%
%
%
%
%
%
%
%
%


\bibitem{Gibs_Boltz_entrpy_FQM} S. Goldstein,  J. L. Lebowitz, R. Tumulka, and N. Zanghi, arXiv:1903.11870v1 [cond-mat.stat-mech] (2019).

\bibitem{StatMechHuang} K. Huang, \emph{Statistical Mechanics} (John Wiley and Sons, 1987).

\bibitem{UniasPresrvEntropyUSen} F. Hulpke, U. V. Poulsen, A. Sanpera, A. Sen(De), U. Sen, and M. Lewenstein, Found. Phys. \textbf{36}, 477 (2006).


\bibitem{Kac_markovJumpBoltzGibsEntrpy} M. Kac, in \emph{Foundations of kinetic theory}, pp 171-197, J. Neyman (editor),
Proceedings of the Third Berkeley Symposium on Mathematical Statistics and Probability,
vol. III. (Berkeley: University of California Press, 1956).



\bibitem{Bell_spek_unspek_book} J. S. Bell, \emph{Speakable and Unspeakable in Quantum Mechanics}, (Cambridge university press, 1989)

\bibitem{Penrose_large_small_mind} R. Penrose, A. Shimony, N. Cartwright, and S. Hawking, \emph{The Large, the Small and the Human Mind} (Cambridge University Press, 
2000).

\bibitem{CTC_cgargeBenett_Cavalcanti} E. G. Cavalcanti and N. C. Menicucci, arXiv:1004.1219 [quant-ph] (2010).

\bibitem{YouQu17} C. S. Sudheer Kumar, Poster presentation at the ``Young Quantum-2017'' meeting held at 
Harish-Chandra Research Institute, Allahabad, India during 27 February - 01 March 2017 (\url{http://www.hri.res.in/~confqic/youqu17/}). 


\bibitem{Distrbtn_XY_each_Normal} A. Seijas-Macias and A. Oliveira, Probability and Statistics \textbf{32}, 87 (2012).



\end{thebibliography}

\end{document}